\documentclass[preprint,aps,epsfig]{revtex4}
\usepackage{graphicx}

\newcommand{\bscco}{BiSr$_2$Ca$_2$Cu$_2$O$_{8+x}$}
\begin{document}

\title{Long Josephson Junction in a Resonant Cavity}

\date{\today}

\begin{abstract}
We present a model to describe an underdamped long Josephson junction
coupled to a single-mode electromagnetic cavity, and carry out
numerical calculations using this model in various regimes.  The
coupling may occur through either the electric or the magnetic field
of the cavity mode.  When a current is injected into the junction, we
find that the time-averaged voltage exhibits self-induced resonant
steps due to coupling between the current in the junction and the
electric field of the cavity mode. These steps are similar to those
observed and calculated in small Josephson junctions.  When a soliton
is present in the junction (corresponding to a quantum of magnetic
flux parallel to the junction plates), the SIRS's disappear if the
electric field in the cavity is spatially uniform. If the cavity mode
has a spatially varying electric field, there is a strong coupling
between the soliton and the cavity mode. This coupling causes the
soliton to become phase-locked to the cavity mode, and produces
step-like anomalies on the soliton branch of the IV
characteristics. If the coupling is strong enough, the frequency of
the cavity mode is greatly red-shifted from its uncoupled value. We
present simple geometrical arguments which account for this behavior.

\end{abstract}

\author{I. Tornes~\cite{email1}} \author{D. Stroud~\cite{email2}}
\affiliation{Department of Physics, The Ohio State University,Columbus, Ohio 43210}


\maketitle

\section{Introduction}

Josephson junction arrays have now been studied for more than twenty
years\cite{ln}.  They are of interest for many reasons.  On a fundamental
level, many unusual physical phenomena may occur in them (Kosterlitz-Thouless
transitions, and quantum phase transitions from superconductor to insulator).
On an applied level, they have been investigated as possible sources of
coherent microwave radiation, which would occur if all the junctions in the
array were to radiate in phase.

These studies have also been extended to Josephson junctions, and
arrays of junctions, coupled to a resonant cavity.   When such a
system is driven by an applied current, it may exhibit equally
spaced self-induced resonant steps (SIRS's) at voltages equal to
multiples of $\hbar\Omega/2e$, where $\Omega$ is the cavity
resonance frequency\cite{larsen}.   More recently, it was shown that
when an {\em array} of underdamped junctions is placed in a
single-mode cavity, and the array is biased on one of the SIRS's,
the array phase-locks and radiates energy coherently into the
cavity, provided that the number of junctions exceeds a critical
threshold\cite{barbara,vasilic,vasilic1,vasilic2}. The SIRS's are
thought to occur because the cavity acts back on the array like an
a. c. driving field, producing voltage plateaus analogous to Shapiro
steps.   The physics of the threshold is believed to be that, since
each junction is coupled to the {\em same} mode, it is effectively
coupled to all the other junctions.  Thus, there is a long range
interaction for which phase locking occurs when the coupling
strength, multiplied by the junction number, exceeds a critical
value.

Several models have been developed to describe these
steps\cite{filatrella,filatrella1,almaas,filatrella2,bonifacio}.
These models generally succeed in reproducing many of the salient
features of the experiments: SIRS's at the expected voltage, and an
increase in the energy in the cavity proportional to the square of the
number of junctions.  The models have now been extended to
two-dimensional (2D) junction arrays\cite{almaas1}, where they show
that, on a given step, the 2D arrays radiate much more energy into the
cavity than the 1D arrays. However, recent work\cite{filatrella2}
indicates that, in contrast to experiment, the numerical models always
result in the junctions' being phase-locked to each other whenever
they are biased on a SIRS.

These theories have also been extended into the {\em quantum}
regime\cite{shnirman,buisson,blais2,everitt,alsaidi,blais,nori}.  This
regime is relevant when the junctions are very small, so that the
non-commutation of the Cooper pair number operator and the phase
operator is important.  In this regime, the states of the junction (or
junctions) and the cavity become entangled, and uniquely quantum
phenomena such as Rabi oscillations may be detectable in suitable
experiments.  Since such junction-cavity systems may be controllable
externally, they may be useful as qubits in quantum computation.  For
this reason, the quantum regime has lately received extensive
attention.

Several groups have also investigated the IV characteristics of a
Josephson junction coupled to a phonon mode.  This system is
formally analogous to the junction-resonant cavity system, in that
both involve coupling between a junction and a harmonic oscillator
mode. Helm {\it et al.}\cite{helm1,helm2} have developed a model
describing the coupling between longitudinal phonons and intrinsic
Josephson oscillations in cuprate superconductors, which, when
sufficiently anisotropic, behave like stacks of underdamped
Josephson junctions\cite{kleiner,kleiner1,kleiner2}.  These theories
have also been extended into the quantum regime\cite{irish}.

In all the above calculations, each superconducting island was
treated as a {\em small} object, with only two degrees of freedom:
the phase of the superconducting order parameter, and the number of
Cooper pairs on each island.  If the island is still large enough
for Cooper pair number and phase to commute, then these two degrees
of freedom can be treated classically. One then obtains a set of
classical coupled second-order differential equations describing the
dynamics of the junction array.  For small islands, the phase and
number are non-commuting operators, but there are still only two
such operators per island.

But one may often be interested in {\em long} junctions, and arrays of
long junctions.  In such systems, the phase difference across the junction
typically depends on position along the junction.  When such a junction
is coupled to a cavity, its behavior may therefore differ from that of small
junctions.

In this paper, we present a simple model for a {\em long} Josephson
junction coupled to a single-mode resonant cavity.   We present our
model in the classical regime, in which number and phase degrees of
freedom commute. However, the model should be readily extendible
into the quantum regime. We consider both spatially uniform and
spatially varying coupling between the junction and the cavity. If
the coupling is uniform, we find that, as in the case of small
junctions, there are SIRS's in the IV characteristics provided that
there is no magnetic field parallel to the junction plates. But when
such a magnetic field is introduced (in the form of a soliton
propagating in the junction), the SIRS's are no longer present for
{\em uniform} coupling. We will give simple arguments why the steps
are not to be expected in this case.  If the coupling is {\em
non-uniform}, however, we find that a moving soliton in the long
junction does couple to the cavity mode.  This coupling perturbs the
current-voltage (IV) characteristics of the junction, producing
step-like structures in the IV characteristics, and excitation of
the cavity mode.  If the coupling is strong enough, this interaction
produces a substantial shift in the frequency of the cavity mode.
All these perturbations and steps can also be understood by simple
physical arguments.

As in the case of short junctions, some previous work has also been
done treating the coupling of intrinsic long junctions to phonons.  In
particular, Preis {\it et al.}\cite{preis} have treated this coupling
in systems with a magnetic field parallel to the junction plates.
They also find resonances and phonon frequency shifts analogous to
those found in the present work, although they use a somewhat
different mathematical model.  Besides this work, several other groups
have studied the interaction of solitons in long Josephson junctions
with various types of harmonic modes.  For example,
Kleiner\cite{kleiner3} has investigated the interaction of solitons in
stacks of long junctions with the cavity resonances whose frequency
and wave number are determined by the stack geometry and by the
Josephson plasma frequency. Machida {\it et al.}\cite{machida} have
numerically studied emission of electromagnetic waves by solitons in
stacks of long junctions which are coupled to the Josephson plasma
resonances of the stack. Salerno {\it et al.}\cite{salerno}, and
Gr\/{o}nbech-Jensen\cite{jensen}, have considered the coupling of
solitons in long junctions to externally applied microwave fields.
Below we compare some of these results with our own.

Since SIRS's may occur in long junctions, they may be useful as
sources of sub-THz radiation.   Hence, {\em stacks} of long
junctions may be even more useful as such sources. Since the most
anisotropic high-T$_c$ cuprate superconductors behave as stacks of
long underdamped Josephson junctions, these natural materials, too,
may serve as a source of coherent sub-THz radiation\cite{tornes}.

The remainder of this paper is arranged as follows.  In Section II, we
derive the equations of motion for treating a long junction coupled to
a single-mode resonant cavity.  In Section III, we present numerical
results obtained from these equations of motion, for both no-soliton
and soliton initial conditions, and for both uniform and non-uniform
junction-cavity coupling.  Finally, in Section IV, we discuss the
numerical results and their experimental implications, and also
suggest some possible extensions of this work.  A detailed derivation
of the Lagrangian for the junction-cavity interaction is presented in
an Appendix.

\section{Derivation of Equations of Motion.}

\subsection{Lagrangian for a Long Josephson Junction}

We will consider a long Josephson junction characterized by a
gauge-invariant phase difference $\phi(x, y, t)$, where $x$ and $y$
are spatial coordinates and $t$ is the time.  $\phi(x, y, t)$
represents the phase difference between the two plates of the junction
at points $(x, y)$.  We assume that the junction has length $L_x$ in
the $x$ direction and $L_y \gg L_x$ in the $y$ direction.  For
mathematical convenience, we assume periodic boundary conditions in
the $x$ direction, so that $\phi(x + L_x, y, t) = \phi(x, y, t)$.  We
also assume that the phase varies only in the $x$ direction.

In the absence of the cavity, the junction Lagrangian is $L_{junc} =
\int dx {\cal L}_{junc} L_y$.  The Lagrangian density ${\cal L}_{junc}
= K - U$, where $K$ and $U$ are the kinetic and potential energies per
unit area.  $K = \epsilon d E^2/(8\pi)$ is the energy density stored
in the cavity electric field, $d$ is the spacing between the junction
plates, and $\epsilon$ is the dielectric constant of the material
within the junction.

With the help of the Josephson relation $\phi_t = \frac{2eV}{\hbar} =
\frac{2edE}{\hbar}$, where $V$ is the voltage drop across the
junction, $e$ is the magnitude of the electronic charge, and the
subscript denotes a time derivative, this becomes $K =
\frac{\epsilon\hbar^2\phi_t^2}{8\pi(2e)^2d}$.

The potential energy density $U = U_J + U_B$.  $U_J = -\frac{\hbar
J_c}{2e}\cos\phi$ is the Josephson coupling energy density,  $J_c$
being the critical current density.  The magnetic field energy per
unit junction area is
\begin{equation}
U_B =\frac{dB^2}{8\pi}, \label{eq:ub}
\end{equation}
where ${\bf B} = {\bf \nabla}\times {\bf A}$ is the local magnetic
induction. The vector potential ${\bf A}$ is related to the
gauge-invariant phase $\phi$ by $\phi(x,y) = \theta(x,y) -
\frac{2\pi}{\Phi_0}\int{\bf A}(x,y)\cdot{dl}$,  where $\theta$ is
the local phase difference across the junction in a particular
gauge, $\Phi_0 = hc/(2e)$ is the flux quantum, and the second term
is the integral of the vector potential across the junction at the
point $(x, y)$. We now assume that ${\bf B} = B{\bf \hat{y}}$, i.
e., is parallel to the plates of the junction. If we choose the
gauge ${\bf A} = -A(x){\bf \hat{z}}$, then $\phi(x,y)=\theta(x,y) +
\frac{2\pi}{\Phi_0}A(x)d$ and therefore
\begin{equation}
B = \frac{\Phi_0}{2\pi d}(\phi_x - \theta_x),
\label{eq:b}
\end{equation}
where the subscript denotes a partial derivative with respect to $x$.

If the superconductor in the upper plate has a complex order
parameter $\Psi_u(x,y) = |\Psi_u|\exp(i\theta_u)$, then the
corresponding in-plane current density may be written
\begin{equation}
 {\bf J}_u(x,y) =
\frac{e^*}{2m^*}\left[|\Psi_u|^2{\bf \nabla}\theta_u -
\frac{(e^*)^2}{m^*c}{\bf A}|\Psi_u|^2\right], \label{eq:curr}
\end{equation}
where $e^* = 2e$, $m^*$ is the effective mass of a Cooper pair, and
the gradient is taken in the xy plane.  The coefficient of {\bf A}
in eq.\ (\ref{eq:curr}) can be identified with the London
penetration depth $\lambda$ of the material in the upper plate, via
the relation $4\pi (e^*)^2 |\Psi_u|^2/(m^* c^2) = 1/\lambda^2$.
Similar equations hold for $\theta_\ell$ on the lower plate.  Hence
${\bf \nabla}\theta \equiv {\bf \nabla}(\theta_u - \theta_\ell)$
satisfies
\begin{equation}
{\bf \nabla}\theta = \frac{4\pi (e^*)^2\lambda^2}{\hbar c^2}({\bf J}_u -
{\bf J}_\ell),
\label{eq:gradth}
\end{equation}
where we have assumed that materials in the upper and lower plates
have the same penetration depth.  Substituting eq.\
(\ref{eq:gradth}) into eqs.\ (\ref{eq:b}) and (\ref{eq:ub}), we
obtain
\begin{equation}
U_B = \frac{d}{8\pi}\left[\frac{\Phi_0\phi_x}{2\pi d} - \frac{4\pi
\lambda^2}{dc}(J_u - J_\ell)\right]^2,
\end{equation}
where we have assumed that $\phi$ varies only in the $x$ direction, as is
reasonable for ${\bf B}\| {\bf \hat{y}}$.

Given ${\cal L}_{junc}$, the equations of motion can be obtained from\cite{fw}
\begin{equation}
\frac{\partial}{\partial t}\frac{\partial {\cal L}_{junc}}{\partial\phi_t} +
\frac{\partial}{\partial x}\frac{\partial {\cal L}_{junc}}{\partial\phi_x}
-\frac{\partial {\cal L}_{junc}}{\partial\phi} = 0.
\label{eq:ljunc}
\end{equation}
In obtaining these equations and henceforth, we assume $J_u - J_\ell
= 0$.  The equations then take the form
\begin{equation}
\phi_{xx} - \frac{1}{\bar{c}^2}\phi_{tt} - \frac{1}{\lambda_J^2}\sin\phi = 0
\end{equation}
where $\bar{c}^2 = c^2/\epsilon$ and $1/\lambda_J^2 = 4\pi e^* d
J_c/(\hbar c^2)$ is the squared Josephson penetration depth. This is
the well-known equation for the phase in a long Josephson junction,
in the limit of no dissipation.  It leads to many solutions,
including the sine-Gordon soliton~\cite{ustinov} $\phi(x, t) =4
\tan^{-1}\left[\exp\frac{(x-x_0)/\lambda_J - \beta t}{\sqrt{1 -
\beta^2}}\right]$ where $\beta = v/\bar{c}$ is a scaled velocity.
This corresponds to an excitation traveling in the positive $x$
direction with speed $v$.  This excitation carries a single quantum,
$\Phi_0 = hc/2e$ of magnetic flux in the positive $z$ direction. The
soliton is relativistically Lorentz-contracted and cannot exceed a
speed of $\bar{c}$.  A similar antisoliton travels with constant
speed $v$ in the negative $x$ direction.

\subsection{Cavity Lagrangian}

We assume that the cavity supports one harmonic oscillator mode,
described by a ``displacement'' variable $q_r$ and its time
derivative $\dot{q}_r$.  A suitable Lagrangian for this mode is
$L_{osc} = \frac{M\dot{q}_r^2}{2} - \frac{Kq_r^2}{2}$, where $M$ is
the ``mass'' of the oscillator mode and $K$ is the ``spring
constant.''  The corresponding Lagrange equation of motion is
\begin{equation}
\frac{d}{dt}\left(\frac{\partial L_{osc}}{\partial \dot{q}_r}\right) -
\frac{\partial L_{osc}}{\partial q_r} = 0,
\label{eq:losc}
\end{equation}
which gives $\ddot{q}_r + \Omega^2 q_r = 0$,  where $\Omega =
(K/M)^{1/2}$ is the oscillator frequency.  Here, we envision the
oscillator as an electromagnetic mode of a suitable resonant cavity.
In this case, for some types of electromagnetic modes, $q_r$ is
proportional to the electric field of the mode.  The formalism
described here would, however, also apply to suitable single-mode
mechanical oscillators.  With some modifications, it would also
apply to electromagnetic modes in which $q_r$ represents the
magnetic, rather than the electric field.

\subsection{Cavity-Oscillator Coupling and Bias Current}

We assume a capacitive coupling between the junction and the
oscillator of the form as is shown in the Appendix, the coupling
Lagrangian takes the form
\begin{equation}
L_{coup} = -\dot{q}_r\int g_E(x)\phi_t(x)dxL_y \equiv  \int {\cal
L}_{coup}dxL_y. \label{eq:lcoup}
\end{equation}
This is a natural extension of coupling assumed in Ref.~\cite{almaas}
to a long Josephson junction [see, in particular, eq.\ (38) of that
paper]. If the cavity electric field is {\em non-uniform}, the
coupling depends on position along the cavity.  In the numerical
examples given below, we consider both position-independent and
position-dependent coupling.

Besides the capacitive coupling, there could, in principle, also be an
{\em inductive} coupling between the junction and the cavity mode.  In
the present work, we do not include this term.  However, if the cavity
electric field has a non-zero curl, there is a corresponding magnetic
field which is already included in eq.\ (\ref{eq:lcoup}).  We include
this type of electric field in some of our calculations below.  A more
general derivation of the junction-cavity coupling can be found in the
appendix.

We also need to include a term in the Lagrangian corresponding to
the bias current.  This takes the form $ L_{curr} = \frac{\hbar
J_z}{2e}\int dx \phi(x) L_y$.

\subsection{Equations of Motion}

The equations of motion are obtained from the analogs of eqs.\
(\ref{eq:ljunc}) and (\ref{eq:losc}), but using the full Lagrangian.
We derive the equations of motion assuming a sinusoidal coupling
$g_E(x) = g_E\sin(k\,x)$, where $g_E$ is a constant. The equations of
motion for {\em uniform} coupling are discussed below.

The total Lagrangian is
\begin{equation}
L_{tot} =  \int \left({\cal L}_{junc} + {\cal L}_{coup}\right)dx L_y
+ L_{osc} + L_{curr}. \label{eq:ltot}
\end{equation}
The Lagrange equations of motion take the form
\begin{eqnarray}
\frac{\partial}{\partial t}
\left[\frac{\partial}{\partial \phi_t}\left({\cal L}_{junc}
+{\cal L}_{coup}\right)\right] & + &\frac{\partial}{\partial x}
\left[\frac{\partial}{\partial \phi_x}\left({\cal L}_{junc} +
{\cal L}_{coup}\right)\right] - \nonumber \\
&-& \frac{\partial}{\partial \phi}\left({\cal L}_{junc} +
{\cal L}_{coup}\right)
= 0
\label{eq:ljunc1}
\end{eqnarray}
and
\begin{equation}
\frac{d}{dt}\left(\frac{\partial L_{tot}}{\partial \dot{q}_r}\right)
- \frac{\partial L_{tot}}{\partial q_r} = 0.
\label{eq:losc1}
\end{equation}
Carrying out these operations, we obtain the equations of motion as
\begin{equation}
\phi_{xx} - \frac{1}{\bar{c}^2}\phi_{tt}-\frac{1}{\lambda_J^2}
\left(\sin\phi -\frac{J_z}{J_c}\right) +g^\prime\,\sin(k\,x) \ddot{q}_r = 0,
\label{eq:phixx}
\end{equation}
where
\begin{equation}
g^\prime = \frac{4\pi d (e^*)^2}{\hbar^2c^2} g_E,
\end{equation}
and
\begin{equation}
\ddot{q}_r + \Omega^2 q_r =
\frac{\hbar^2 c^2}{4\pi d (e^*)^2M}g^\prime \int \,\sin(k\,x)\phi_{tt}dxL_y.
\label{eq:ddotq}
\end{equation}

The above equations of motion are derived for a cavity electric
field which varies sinusoidally with position.  To obtain the
equations of motion for a {\em uniform} cavity electric field, one
simply replaces the term $g^\prime\sin(kx)$ in eqs.\
(\ref{eq:phixx}) and (\ref{eq:ddotq}) (and subsequent equations) by
$g^\prime$.

Eqs.\ (\ref{eq:phixx}) and (\ref{eq:ddotq}) do not include any
damping.  We simply incorporate damping by hand, by adding the
appropriate terms to these equations.  The resulting equations take
the form
\begin{equation}
\phi_{xx} - \frac{1}{\bar{c}^2}\phi_{tt} -\frac{\omega_p}{Q_J\bar{c}^2}\phi_t
- \frac{\omega_p^2}{\bar{c}^2}
\left(\sin\phi - \frac{J_z}{J_c}\right) + g^\prime\,\sin(k\,x)\ddot{q}_r = 0,
\label{eq:qj}
\end{equation}
and
\begin{equation}
\ddot{q}_r + \frac{\Omega}{Q_c}\dot{q}_r + \Omega^2 q_r
= \frac{\hbar^2 c^2}{4\pi d (e^*)^2M}g^\prime \int \,\sin(k\,x)\phi_{tt} dxL_y.
\label{eq:qc}
\end{equation}
Here we have introduced dimensionless junction and cavity quality
factors $Q_J$ and $Q_c$, and a Josephson plasma frequency $\omega_p
= \bar{c}/\lambda_J = [4\pi e^* dJ_c/(\hbar\epsilon)]^{1/2}$.  In
terms of the junction parameters, $Q_J = \epsilon\omega_p/\sigma$,
where $\sigma$ is the conductivity of the medium within the
junction. The additional terms in eqs.\ (\ref{eq:qj}) and
(\ref{eq:qc}) insure that, in the absence of cavity-junction
coupling, the equations of motion reduce to the standard results for
a long Josephson junction with damping\cite{ustinov}, and for a
damped harmonic oscillator.

\subsection{Reduction to a Set of Coupled First-Order Equations}

Eqs.\ (\ref{eq:qj}) and (\ref{eq:qc}) are conveniently solved
numerically if they are converted into a set of coupled first-order
differential equations.  Thus, we introduce the momenta canonically
conjugate to $q_r$ and $\phi$, namely $p_r = M\dot{q}_r
-E_0g^\prime\int\,\sin(k\,x)\phi_t dx L_y$ and $p_\phi =
E_0\phi_t/\bar{c}^2 - g^\prime \,\sin(k\,x)E_0\dot{q}_r$, where $E_0
= \frac{\hbar^2\,c^2}{4\pi d (e^*)^2}$.  Substituting these
variables into eqs.\ (\ref{eq:qj}) and (\ref{eq:qc}), and
rearranging, we obtain the following set of four coupled partial
differential equations:
\begin{eqnarray}
\dot{q}_r & = &\frac{1}{K}
\left(\frac{p_r}{M} + \frac{\bar{c}^2 g^\prime}{M}\int \,\sin(k\,x)p_\phi dx
L_y\right) \nonumber \\
\dot{p}_r & = & -M\Omega^2 q_r - \frac{M\Omega}{Q_c}\dot{q}_r \nonumber \\
\phi_t & = & \frac{\bar{c}^2}{E_0}p_\phi + g^\prime \,\sin(k\,x)\bar{c}^2\dot{q}_r \nonumber \\
(p_\phi)_t & =&  E_0\phi_{xx}-\frac{E_0\omega_p^2}{\bar{c}^2}
\left(\sin\phi - \frac{J_z}{J_c}\right) - \nonumber \\
& - &\frac{\omega_p\,E_0}{\bar{c}^2}
\frac{\phi_t}{Q_J},
\end{eqnarray}
where $K = 1 - \frac{E_0\bar{c}^2 {g^\prime}^2L_x}{M} \int \, \sin^2(k\,x) dx L_y$

\subsection{Equations of Motion in Dimensionless Form}

It is convenient to rewrite these equations using a dimensionless time
$\tau = \omega_pt$, length $\xi = x/\lambda_J$, and cavity frequency
$\tilde{\Omega} = \Omega/\omega_p$.  We also introduce the
dimensionless variables $\tilde{p}_r = p_r/(M\bar{c})$,
$\tilde{p}_\phi = (M\bar{c}^2/E_0)^2(p_\phi/(M\omega_p)$.
$\tilde{E}_0 = E_0/(M\bar{c}^2)$, $\tilde{q}_r = q_r/\lambda_J$, and
finally $\tilde{g} = g^\prime L_y\bar{c}^2$.  Finally, we require that
$k = 2\pi m/L_x$, where $m$ is an integer, consistent with periodic
boundary conditions.  With these substitutions, and further
rearrangement, the equations of motion become
\begin{eqnarray}
\dot{\tilde{q}}_r & = &\frac{1}{\tilde{K}} \left(\tilde{p}_r +
\tilde{E}_0^2\tilde{g} \int \tilde{p}_\phi(\xi_x)\,\sin(\frac{2\pi
m\xi\lambda_J}{L_x})d\xi_x\right) \nonumber \\
\label{eq:f1} \nonumber \\
\dot{\phi}&  = & \tilde{E}_0\tilde{p}_\phi +
\tilde{g}\frac{\lambda_J}{L_y}\dot{\tilde{q}}_r
\sin(\frac{2\pi m\xi\lambda_J}{L_x}) \label{eq:f2} \nonumber \\
\tilde{E}_0\dot{\tilde{p}}_{\phi} &  = &
\phi_{\xi\xi} - \left(\sin\phi - \frac{J_z}{J_c}\right) -
\frac{\dot{\phi}}{Q_J}
\label{eq:f3} \nonumber \\
\dot{\tilde{p}}_r & = & - \frac{\tilde{\Omega}}{Q_c}\dot{\tilde{q}}_r -
\tilde{\Omega}^2\tilde{q}_r,
\label{eq:f4}
\end{eqnarray}
where the dot is a derivative with respect to $\tau$ and $\tilde{K}
= 1-\frac{\tilde{E}_o (\tilde{g})^2}{L_y} \int \sin^2(k\,x)\,dx $.

\section{Numerical Results}

\subsection{Algorithm}

We solve eqs.\ (\ref{eq:f4}) numerically by discretizing them on a
spatial scale of $\Delta$. Thus, $\phi_{\xi\xi} \rightarrow
\frac{\phi_{j+1} - 2\phi_j + \phi_{j-1}}{\Delta^2}$,  where $j = 1,
..., N$, and $N = \frac{L_x}{\Delta\,\lambda_J}$ is the number of
discrete sections of the long junction. In discretized form, eqs.\
(\ref{eq:f4}) become
\begin{eqnarray}
\dot{\tilde{q}}_{r}& = & \frac{1}{\tilde{K}}\left[\tilde{p}_{r} +
\Delta\tilde{E}_0^2\tilde{g}
\sum_{i=1}^N \tilde{p}_{\phi_i}\,\sin(\frac{2\pi m i}{N})\right] \label{eq:f1p} \nonumber \\
\dot{\phi}_i & = & \tilde{E}_0\tilde{p}_{\phi_i}
+ \tilde{g}\frac{\lambda_J}{L_y}\,\sin(\frac{2\pi m i}{N})\dot{\tilde{q}}_{r} \label{eq:f2p} \nonumber \\
\tilde{E}_0\dot{\tilde{p}}_{\phi_i} & = &
\frac{\phi_{i+1} - 2\phi_i + \phi_{i-1}}{\Delta^2} - \left(\sin\phi_i
- \frac{J_z}{J_c}\right) \nonumber \\
&-& \frac{\dot{\phi}_i}{Q_J} \nonumber \label{eq:f3p} \\
\dot{\tilde{p}}_{r} & = & -\frac{\tilde{\Omega}}{Q_c}\dot{\tilde{q}}_{r}
- \tilde{\Omega}^2\tilde{q}_{r}. \label{eq:f4p}
\end{eqnarray}
With this discretization, we are basically treating the long
junction as N inductively coupled small junctions, each of which has
critical current $I_c = J_c (L_xL_y)$.

We have solved eqs.\ (\ref{eq:f4p}) numerically
using a constant-time-step fourth-order Runge-Kutta method with a
time step $\Delta\tau = 0.001$.  We begin the simulation by
initializing the variables $\tilde{p}_r$, $\tilde{q}_r$,
$\tilde{p}_{\phi_j}$ and the parameter $\frac{J_z}{J_c}$ to zero; we
have made various choices for initial values of $\phi_j$ as
discussed below.  For a given $J_z/J_c$, we integrate the
differential equations from $\tau = 0$ to $\tau = 5 \times 10^3$,
then evaluate the voltages by averaging over the last $\tau = 2
\times 10^3$ units of time.  The ratio $\frac{J_z}{J_c}$ is then
increased or decreased by 0.01 and the set of equations is solved
again. In all cases, we use periodic boundary conditions in the x
direction.

\begin{figure}
\begin{center}
\includegraphics[width = 0.45\textwidth]{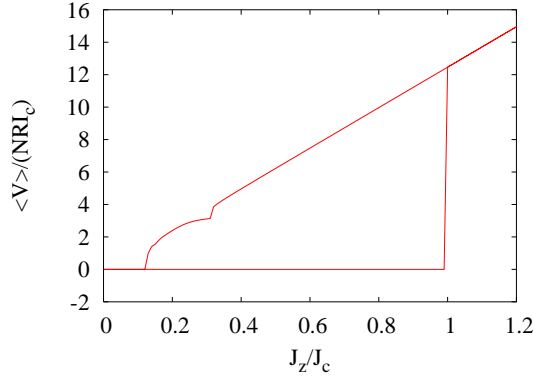}
\caption
{ \small IV curve for single long junction, with $Q_J = 10.0$, $\Delta
= 0.05$, $\tilde{\Omega} = 2.5$, $\tilde{E}_0^2\Delta\tilde{g} = 1.0
\times 10^{-4}$, $Q_c = 10$ and all the phases initialized to zero. A
SIRS is located at $<V>/(NRI_c) \sim \pi$, where $I_c$ is the critical
current of one of the individual small junctions.}
\label{nosolitonIV}
\end{center}
\end{figure}

\begin{figure}
\begin{center}$\begin{array}{cc}
\includegraphics[width = 0.225\textwidth]{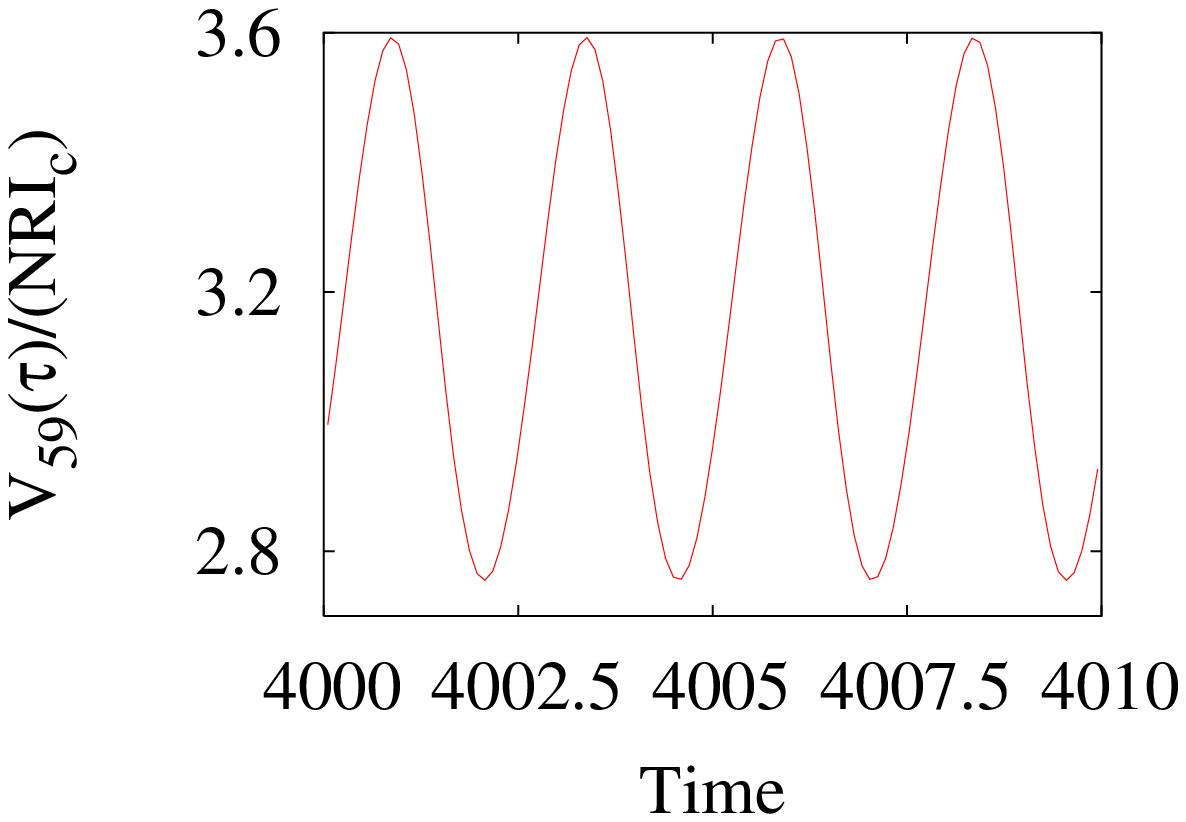} &
\includegraphics[width = 0.225\textwidth]{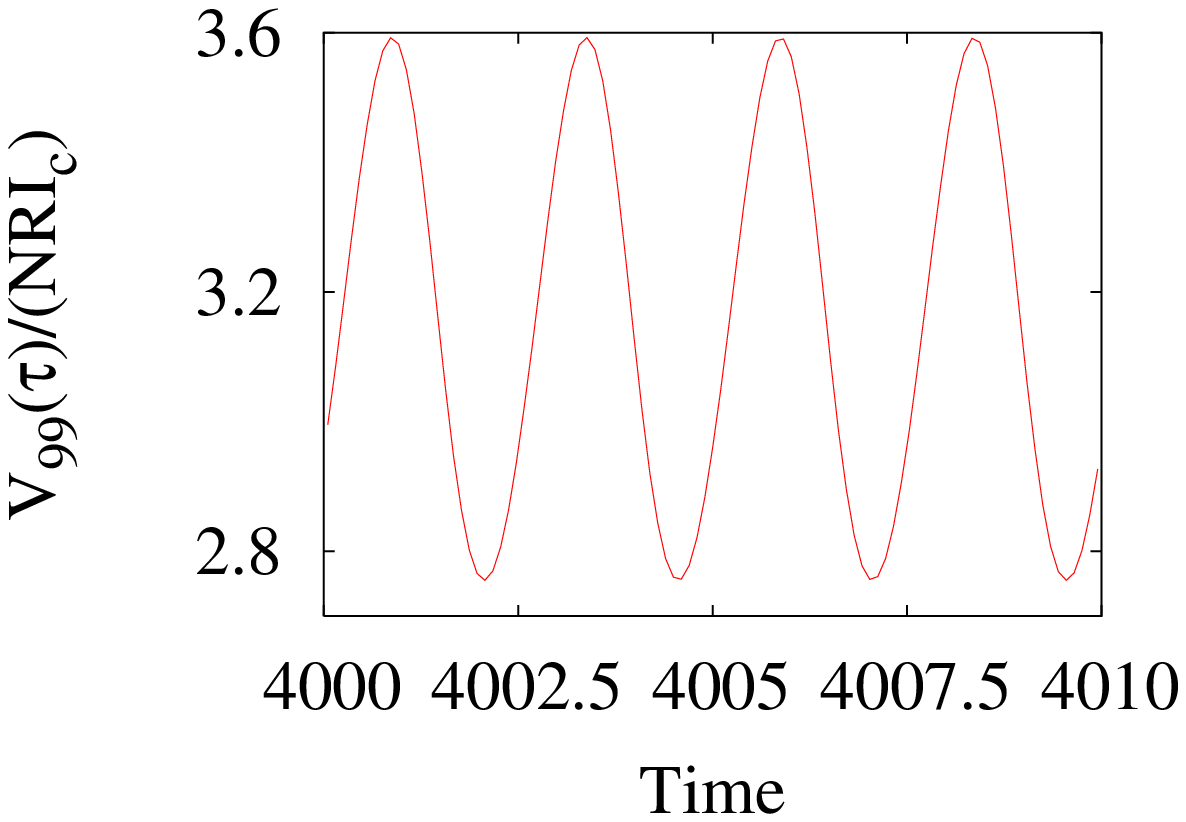} \\
(a) &
(b)
\end{array}$
\caption
{\small $V(\tau)/(NRI_c)$  versus dimensionless time $\tau$
for small junctions no.\ (a) 59 and (b)
99 for $4000 \leq \tau \leq 4010$, plotted
for a voltage on the SIRS of
Fig.\ \ref{nosolitonIV}.}
\label{nosolitonphaseplots}
\end{center}
\end{figure}

\begin{figure}
\begin{center}
\includegraphics[width = 0.45\textwidth]{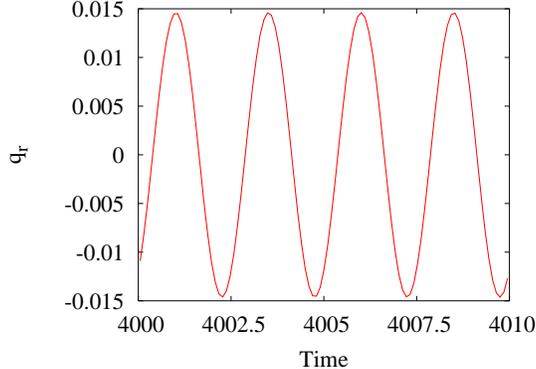}
\caption
{\small Cavity variable $q_r(\tau)$
for the long junction in Fig.\ \ref{nosolitonIV},
at an applied current on the SIRS, plotted versus time $\tau$ for the
time interval $4000 \leq \tau \leq 4010$.}
\label{nosolitoncavityplot}
\end{center}
\end{figure}

We have considered two different initial phase distributions.  To
model a junction containing no soliton, we choose all the initial
phases $\phi_j = 0$ for $j = 1,...,N$.  To model a junction
containing a single soliton, we assume an initial configuration
$\phi_n(\tau = 0) = 2(\frac{\pi}{2}+\tanh{\frac{2\,\pi\,n}{N}})$,
for $n = 1,...,N$.  As has been discussed in Ref.\ \cite{ustinov},
for example, these boundary conditions are consistent with the
presence of a single soliton, which carries one flux quantum.  Our
numerical results, as presented below, do show evidence for a
soliton with these boundary conditions.

\subsection{Results with No Soliton}

In Fig.\ \ref{nosolitonIV}, we show the IV curve for a single long
junction coupled to a cavity, using the no-soliton initial
conditions (all phases chosen to equal zero).  We choose a {\em
position independent} junction-cavity coupling with a constant
$\tilde{g}\tilde{E}_0^2\Delta = 1.0 \times 10^{-4}$ and $N = 120$.
We plot the time-averaged voltages $\langle V \rangle$ in units of
$NRI_c$, where $I_c = J_c(\,L_x\,L_y)$ is the critical current of a
single small junction.  The IV characteristics have a step at
$\langle V\rangle/(NRI_c) = 4\pi\tilde{\Omega}/Q_J$.  This is a
SIRS, similar to that seen in individual small junctions for a
similar model (see Ref.~\cite{almaas}), and occurs at $\langle V
\rangle/(NRI_c) = \pi$. This is the voltage expected from the model
of Ref.\ \cite{almaas}, which predicts that the SIRS's will occur
when $\frac{<V>}{NRI_c}$ is an integer multiple of
$4\pi\frac{\tilde{\Omega}}{Q_J}$.  The step in Fig.\
\ref{nosolitonIV} occurs at $4\pi \frac{2.5}{10} = \pi$ for $n =1$.
On this step, the phases of all the junctions oscillate coherently
and are locked onto the cavity mode.  To illustrate this coherence,
we show in Figs.\ \ref{nosolitonphaseplots}(a) and
\ref{nosolitonphaseplots}(b) plots of the voltages
$\frac{V_i(\tau)}{NRI_c}$ for small junctions nos. $59$ and $99$
over the same time interval.  The plots show that the two junctions
are indeed oscillating periodically and in phase with one another.
In fact, the $\frac{V_i(\tau)}{NRI_c}$ plots for all the small
junctions are identical on this step.

The phase locking between the junctions and the cavity is
illustrated in Fig.\ \ref{nosolitoncavityplot}.  This Figure shows
that the period of oscillation of  $\tilde{q}_r(\tau)$ is
identical to those of $\frac{V_i(\tau)}{NRI_c}$ shown in Figs.\
\ref{nosolitonphaseplots}; both equal
$\frac{2\,\pi}{\tilde{\Omega}}$.

\subsection{Results with Soliton Present.}

The results shown in Figs.\
\ref{nosolitonIV}-\ref{nosolitoncavityplot} are very similar to
those obtained in Refs~\cite{almaas} and \cite{tornes} for a {\em
single} small junction coupled to a cavity.   However, we see very
different results if there is a soliton initially present in the
junction. In this case, it is important that the equation of
motion for the long junction be discretized on a fine scale. If
$N$ is too small, there are spurious steps in the IV
characteristic produced by locking of the soliton to certain
linear excitations generated purely by the numerical
discretization. These steps, which are numerical artifacts of a
too coarse discretization, are well known in real (and discrete)
Josephson ladders, and have been discussed
extensively\cite{zant,watanabe}. In all our calculations below,
$N$ is sufficiently large to avoid these spurious steps.

\begin{figure}
\begin{center}
\includegraphics[width = 0.45\textwidth]{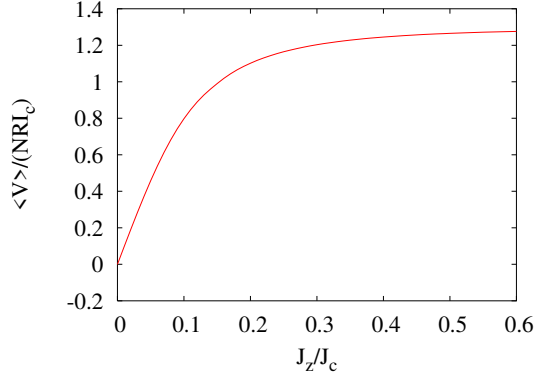}
\caption{\small Current-voltage plot for a single long junction in a
resonant cavity, with $Q_J = 10.0$, $\Delta = 0.05$, $\tilde{\Omega}
= 0.75$, a uniform coupling with $\tilde{g}\tilde{E}_0^2\Delta = 1
\times 10^{-4}$, and $Q_c = 10$.  Our initial conditions are such
that one soliton is present.  This curve is the same for any choice
of the strength parameter $\tilde{g}\tilde{E}_0^2\delta$}
\label{solitonuniform}
\end{center}
\end{figure}

We discuss first a position-independent coupling between the
junction and the cavity.  An example of the calculated
current-voltage characteristics for this case is shown in Fig.\
\ref{solitonuniform}.  The portion of the IV characteristic for
$J/J_c \leq 0.6$ corresponds to a soliton which moves freely through
the long junction.  The soliton behaves like a free, massive, but
relativistic, particle, with limiting velocity $\bar{c}$.  Its
motion is entirely unaffected by coupling to the cavity.  The free
motion of the soliton can be understood in a simple way by the
following argument. If the cavity electric field is {\em uniform},
the coupling between the soliton and the cavity mode is independent
of the soliton position. Thus, the cavity electric field exerts no
force on the soliton, which therefore should still travel with
constant velocity even if there is a strong junction-cavity
coupling, consistent with our numerical results.  In fact, the IV
characteristics shown in Fig.\ \ref{solitonuniform} would be the
same if $\tilde{g} = 0$.

\begin{figure}
\begin{center}
\includegraphics[width = 0.45\textwidth]{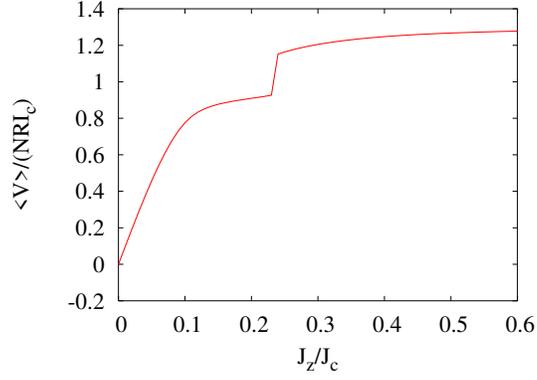}
\caption { \small Current-voltage plot for a single long junction in
a cavity, with $Q_J = 10.0$, $\Delta = 0.05$, $\tilde{\Omega} =
0.75$, $N = 120$, $Q_c = 10$, and coupling
$\tilde{g}\tilde{E}_0^2\Delta \sin(2\pi x/L_x)$, with
$\tilde{g}\tilde{E}_0^2\Delta = 1.\times 10^{-4}$. Our initial
conditions are such that one soliton is present.}
\label{lower_om75_g04}
\end{center}
\end{figure}

\begin{figure}
\begin{center}
\includegraphics[width=0.45\textwidth]{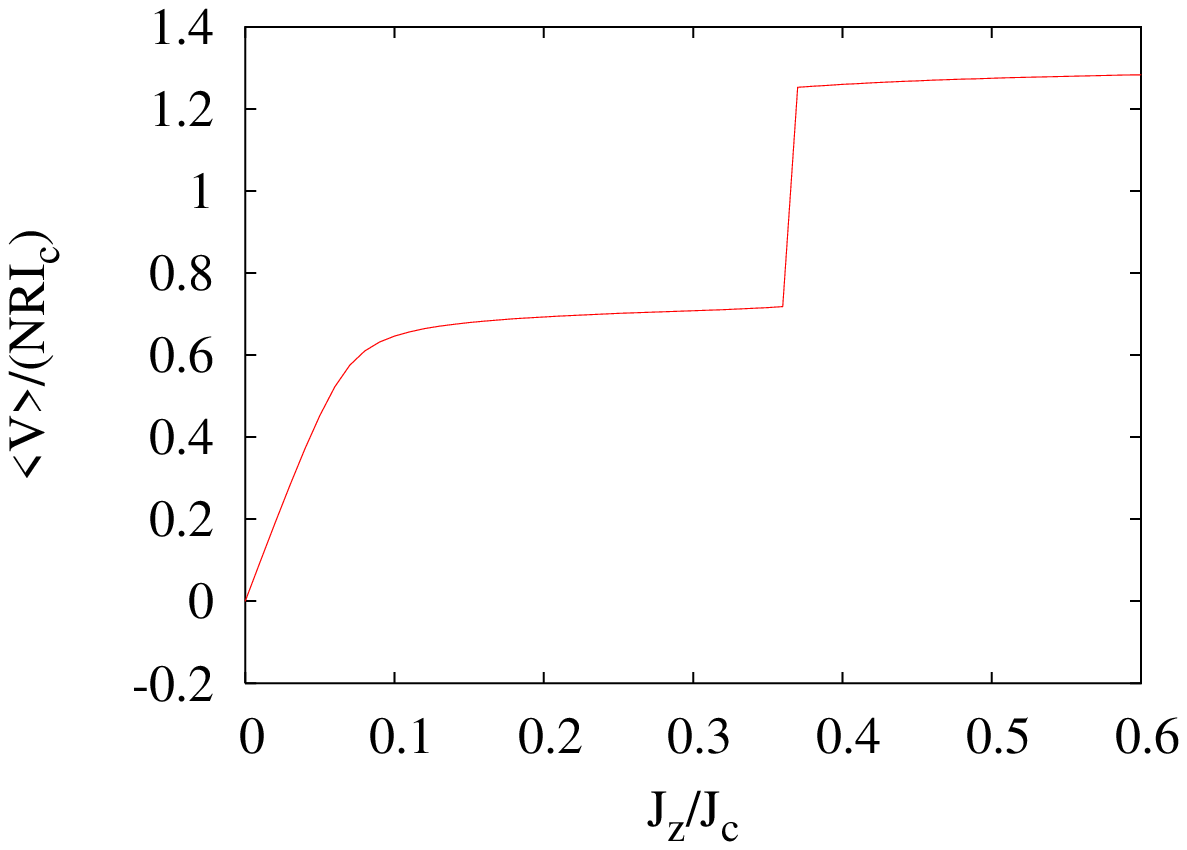} \caption {\small
Same as Fig.\ \ref{lower_om75_g04}, except that
$\tilde{g}\tilde{E}_0^2\Delta = 1.0 \times 10^{-3}$.}
\label{lower_om75_g03}
\end{center}
\end{figure}

\begin{figure}
\begin{center}
\includegraphics[width=0.45\textwidth]{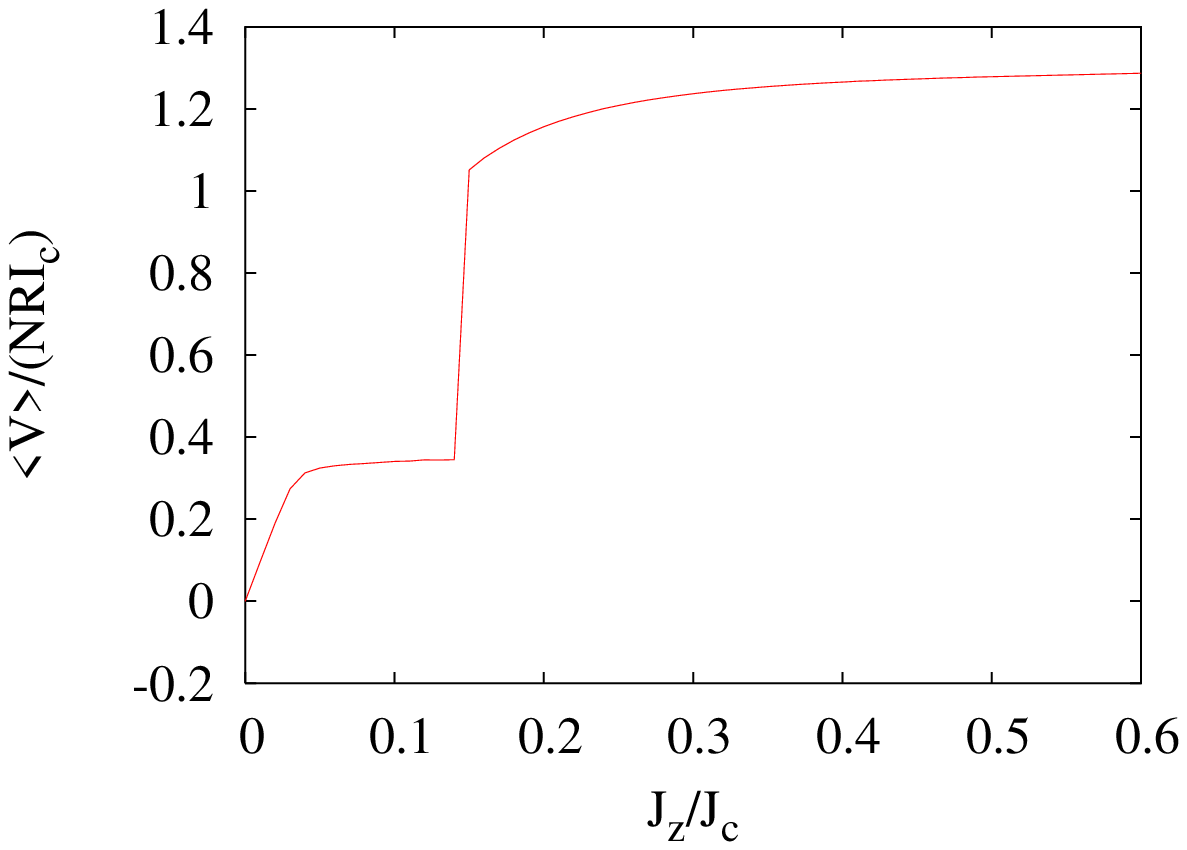} \caption {\small
Same as Fig.\ \ref{lower_om75_g04}, except that
$\tilde{g}\tilde{E}_0^2\Delta = 1.0\times 10^{-2}$.}
\label{lower_om75_g02}
\end{center}
\end{figure}

Next, we consider a {\em position-dependent} junction-cavity
coupling. In this case, the IV characteristics are clearly perturbed
by the cavity.  Figs.\ \ref{lower_om75_g04}-\ref{lower_om75_g02}
shows the soliton branch of the full IV curve for a spatially
varying coupling of the form used in eqs.\ (\ref{eq:qj}) and
(\ref{eq:qc}). By the soliton branch, we mean that part of the IV
curve produced when the current is increased from $J_z/J_c = 0.0$ to
approximately $0.6$. We use the same model parameters as for uniform
coupling, except that we consider several different coupling
strengths, namely $\Delta\tilde{E}_o^2\tilde{g} = g^\prime$, with
$g^\prime = 10^{-4}$, $10^{-3}$, and $10^{-2}$ and the dimensionless
cavity frequency $\tilde{\Omega} = 0.75$. We also take $m = 1$ in
eq.\ (\ref{eq:f1}). There are now clear step-like structures in the
IV characteristics for all three coupling strengths, at $\langle V
\rangle/(NRI_c) \sim 0.95$, $0.65$, and $0.33$ respectively, which
were absent in the case of uniform coupling.  These step-like
structures correspond in each case to the locking of the soliton to
the cavity mode. Specifically, the soliton circulates around the
long junction at a frequency of one cycle per cycle of the cavity
mode.  For the weakest coupling shown, with $g^\prime = 10^{-4}$,
the cavity mode is shifted very little from $\tilde{\Omega} = 0.75$.
The voltage step thus occurs approximately at

\begin{equation} \frac{\langle V \rangle}{NRI_c} =
\frac{4\pi\tilde{\Omega}}{Q_J}, \label{eq:step}
\end{equation}
or $\langle V \rangle/(NRI_c) \sim 0.95$ for these parameters,
corresponding to one cycle of the soliton around the cavity per
unperturbed cavity period.  For the two stronger couplings shown in
Figs.\ \ref{lower_om75_g03} and \ref{lower_om75_g02}, the soliton is
still locked to the cavity, but the coupling is strong enough that
the cavity frequency is shifted substantially down from its
unperturbed value, to about $2\tilde{\Omega}/3$ and
$\tilde{\Omega}/3$ respectively. The corresponding time-averaged
voltage on the step is approximately $\langle V \rangle/(NRI_c)  =
(2/3)[4\pi\tilde{\Omega}/Q_J] \sim 0.66$, and
$(1/3)[4\pi\tilde{\Omega}/Q_J] \sim 0.33$ for these two couplings.

\begin{figure}
\begin{center}
\includegraphics[width=0.45\textwidth]{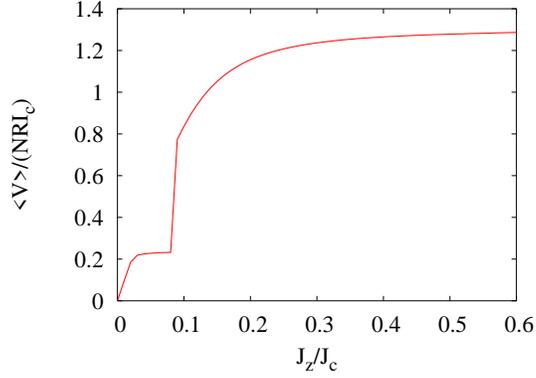} \caption{\small
Same as Fig.\ \ref{lower_om75_g02}, except that $\tilde{\Omega} =
0.5$.}
\label{iv_om5}
\end{center}
\end{figure}

To see the effects of changing the cavity frequency, we have carried
out additional calculations with the same sinusoidal junction-cavity
coupling but different cavity frequencies $\tilde{\Omega}$ and various
coupling strengths.  The voltage plateaus typically vary approximately
linearly with $\tilde{\Omega}$, as predicted by eq.\ (\ref{eq:step}).
An example of this behavior, for the rather large coupling constant
$g^\prime =10^{-2}$, is shown in Fig.\ \ref{iv_om5}. $\langle V
\rangle/(NRI_c)$ is approximately 2/3 the value of
Fig.\ \ref{lower_om75_g02}, as suggested by eq.\ (\ref{eq:step}).

Besides the time-averaged voltages, we have also calculated
time-dependent voltage differences at various points across the
junction for most of the examples shown in
Figs.\ref{lower_om75_g04}-\ref{iv_om5}, and others.  We have also
computed the time-dependent cavity variable $q_r(\tau)$. In all
cases, these calculations provide clear evidence of locking between
the junction and the cavity.

\begin{figure}
\begin{center}
$\begin{array}{cc}
\includegraphics[width=0.225\textwidth]{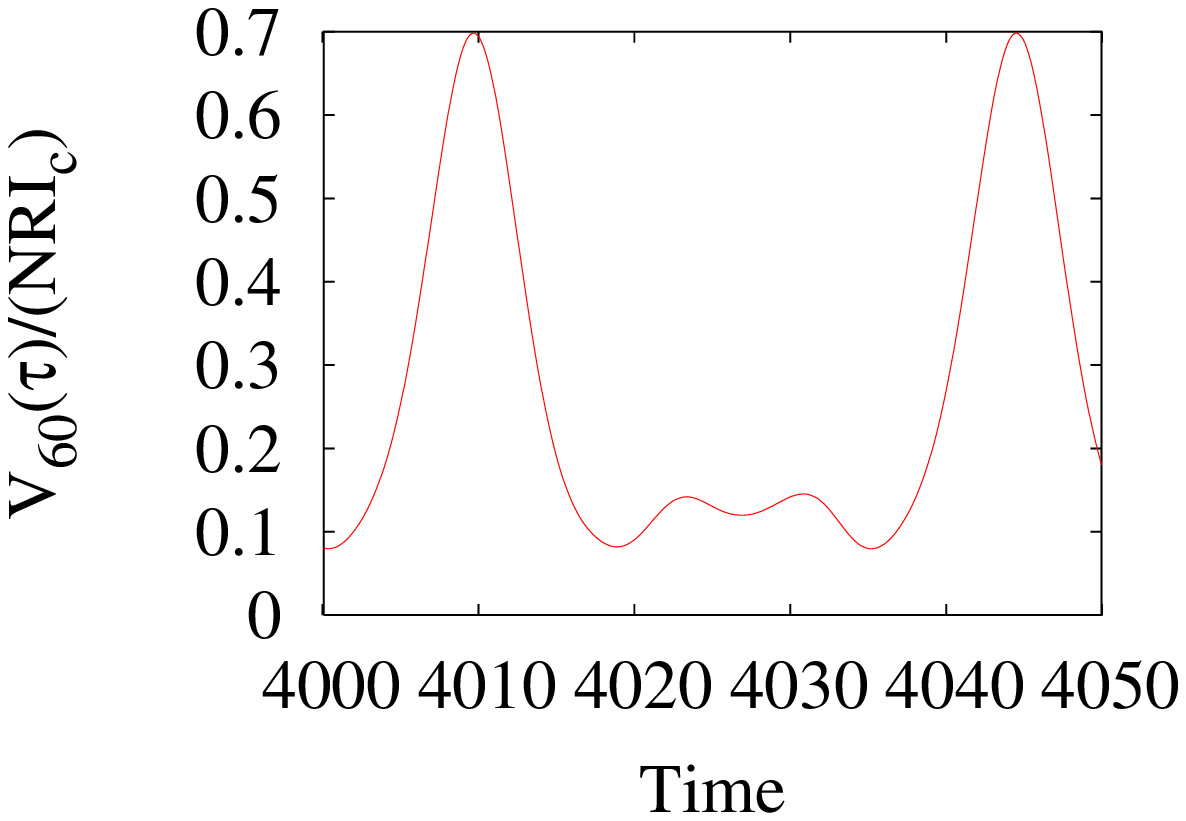} &
\includegraphics[width=0.225\textwidth]{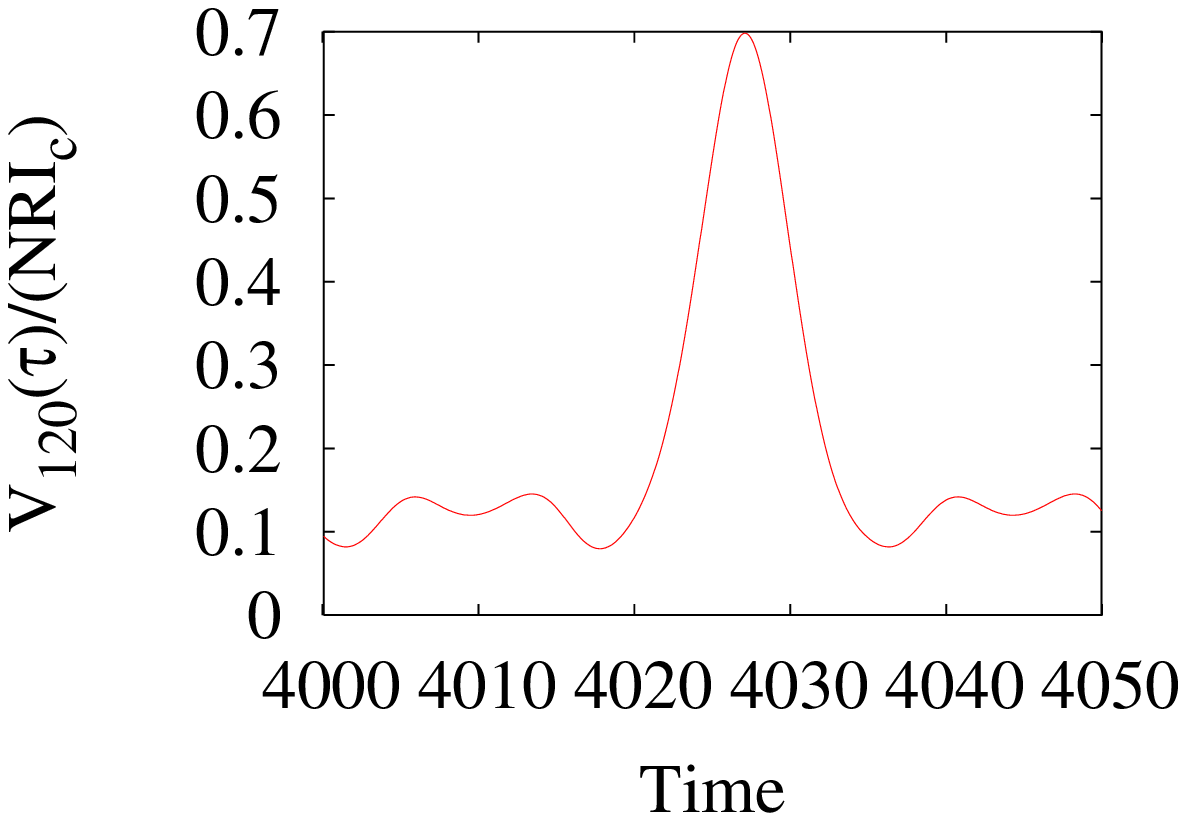} \\
(a) & (b)
\end{array}$
\caption {\small Time-dependent voltage $V(\tau)$ for small
junctions no.\ (a) 60 and (b) 120, for $4000 \leq \tau \leq 4050$,
plotted versus time $\tau$.  Current is such that time-averaged
voltage is on the lowest step-like structure of Fig.\ 8.}
\label{phase60_120}
\end{center}
\end{figure}

\begin{figure}
\begin{center}
$\begin{array}{cc}
\includegraphics[width=0.225\textwidth]{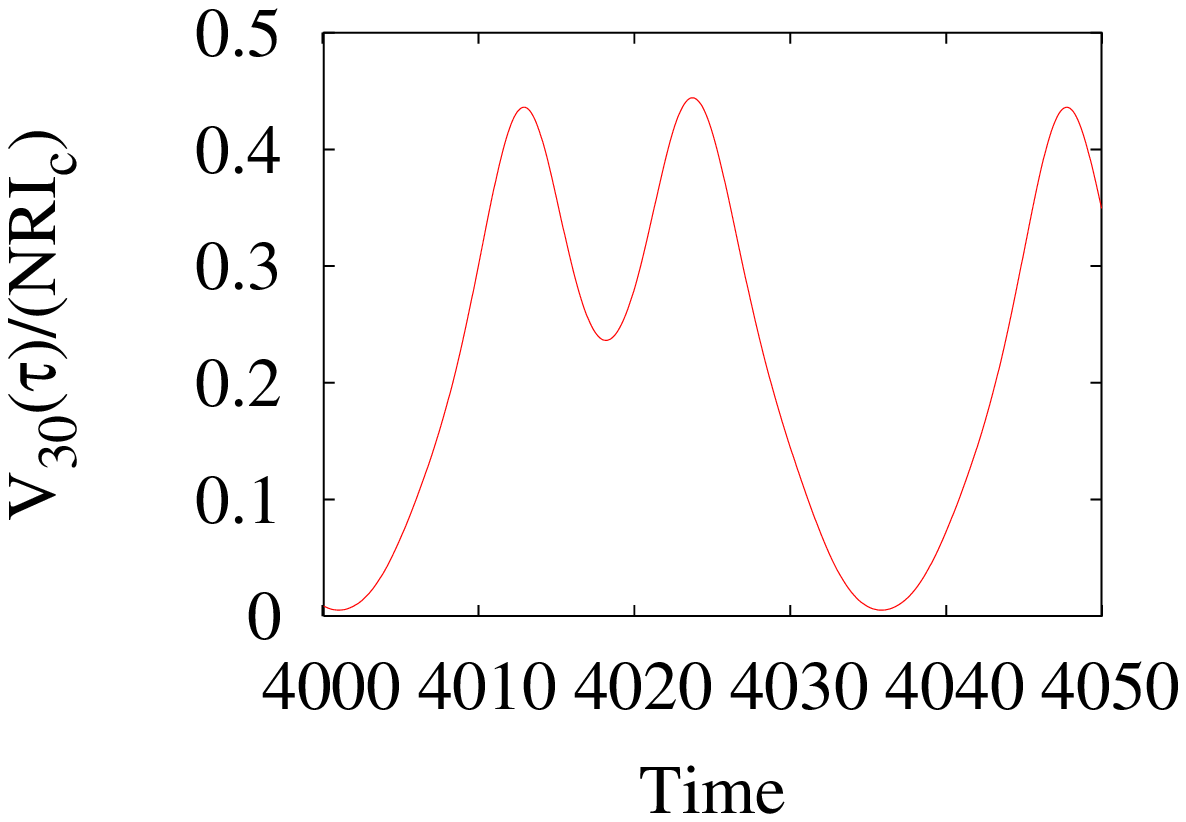} &
\includegraphics[width=0.225\textwidth]{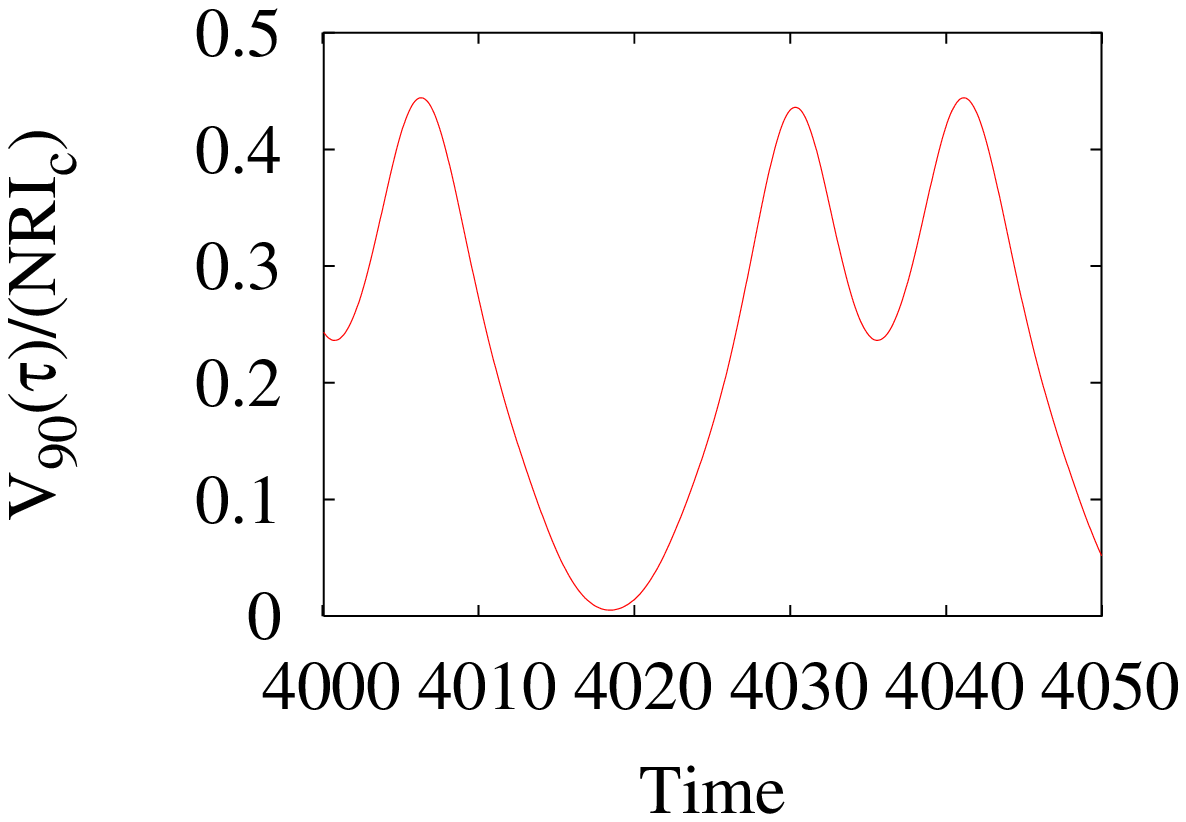}\\
(a) & (b)
\end{array}$
\caption {\small Same as Figs.\ \ref{phase60_120} (a) and (b) except
that we plot voltages for junctions no.\ (a) 30 and (b) 90.}
\label{phase30_90}
\end{center}
\end{figure}

\begin{figure}
\begin{center}
\includegraphics[width=0.45\textwidth]{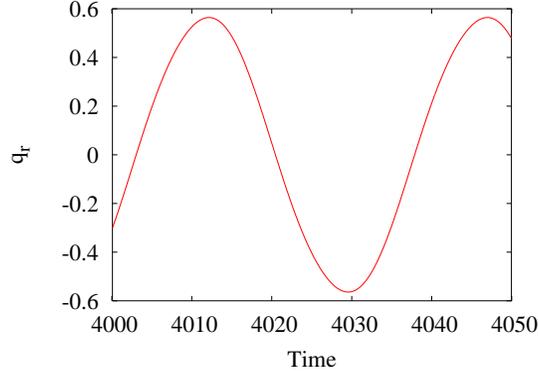}
\caption {\small Cavity variable $q_r(\tau)$ for the long junction
of Fig.\ref{iv_om5}, plotted versus $\tau$ for the voltage on the
lowest step-like structure for $4000 \leq \tau \leq 4050$.}
\label{solitoncavityplot}
\end{center}
\end{figure}

Some representative examples of the voltages are shown in Figs.\
\ref{phase60_120}, \ref{phase30_90}, and \ref{solitoncavityplot}. In
Figs.\ \ref{phase60_120}(a) and (b), and Figs.\ \ref{phase30_90}(a)
and \ref{phase30_90}(b), we show the time-dependent voltages for
mini-junctions nos.\ $60$, $120$, $30$ and $90$ at $\frac{J_z}{J_c}
= 0.04$, for the parameters and frequency of Fig.\ \ref{iv_om5}. The
voltages shown in Figs.\ \ref{phase60_120}(a) and
\ref{phase60_120}(b) have the same characteristic shape and period,
but are $180^o$ out of phase with one another.  The same is true for
Figs.\ \ref{phase30_90} (a) and \ref{phase30_90} (b).  The voltages
of the second pair differ in wave form, but not in period, from
those of the first pair. Indeed, we have found that all $120$
time-dependent voltages on this step have the same period and that
the voltages of all mini-junction pairs separated by exactly
one-half the junction length are identical but 180$^o$ out of phase.
The corresponding behavior of $\tilde{q}_r(\tau)$ is shown in Fig.\
\ref{solitoncavityplot}. $\tilde{q}_r(\tau)$ has the same period as
that of all the individual mini-junction voltages, showing that the
long junction is indeed locked to the cavity mode.  Behavior similar
to that of Figs.\ \ref{phase60_120} and \ref{phase30_90} is also
seen in cases with smaller $g^\prime$.  Because of the weaker
coupling, the amplitude of the cavity parameter $\tilde{q}_r$ on
these steps (not shown) is much smaller than in Fig.\
\ref{solitoncavityplot}. The period of $\tilde{q}_r$ is again the
same as that of all the mini-junction voltages, and the
time-dependent voltages are again identical in pairs, but 180$^o$
out of phase, as in Figs.\ \ref{phase60_120} and \ref{phase30_90}.

We see from Figs.\ \ref{lower_om75_g04}-\ref{iv_om5}, that the
voltage steps are not completely flat.  Despite this slight
curvature, we have confirmed numerically that the voltages of all
the individual mini-junctions have the same period as that of the
cavity.  Furthermore, the time-averaged voltages across each
mini-junction are all the same.   If we examine the voltages at
current values off the step, we find that the cavity and individual
junctions no longer have the same periods.

\begin{figure}
\begin{center}
\includegraphics[width=0.45\textwidth]{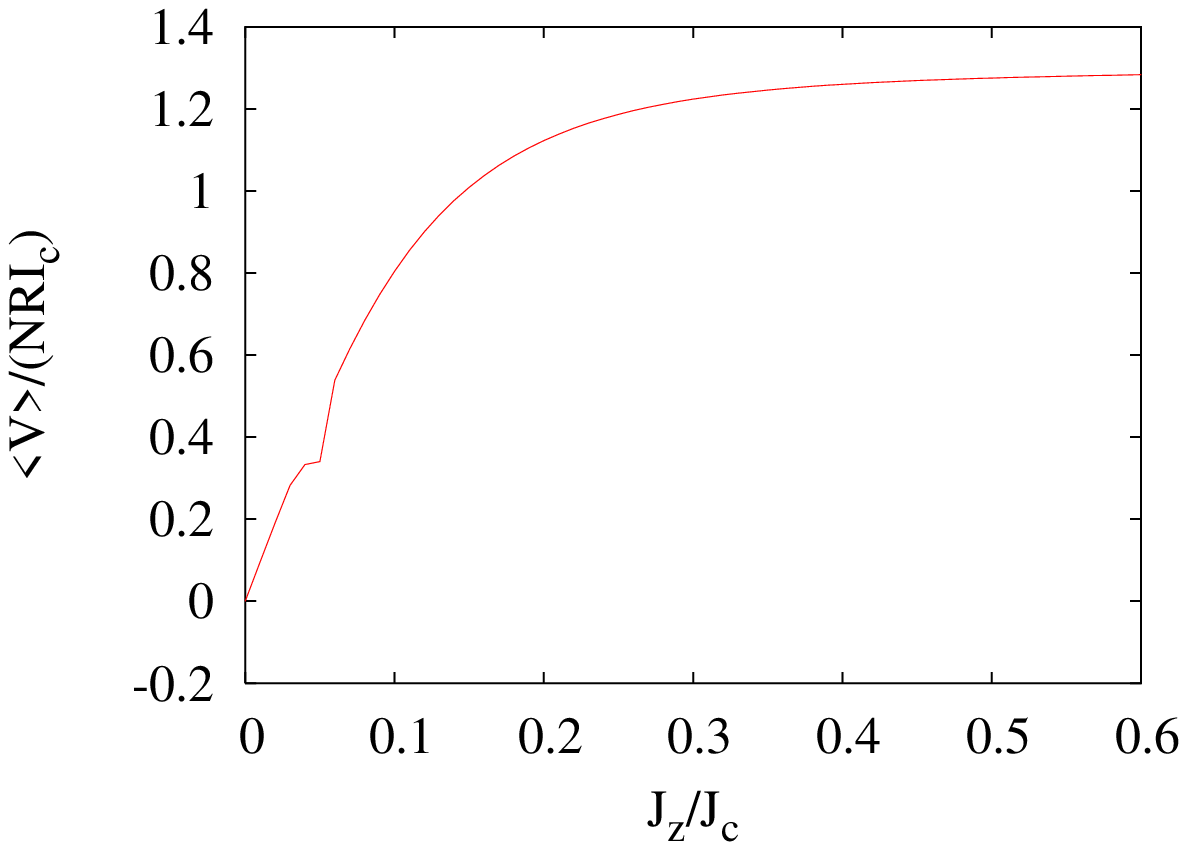}
\caption {\small Current-voltage plot for a single long junction in
a cavity, with same parameters as in Fig.\ \ref{lower_om75_g02},
except that $\tilde{\Omega} = 1.5$ and the coupling is
$\tilde{g}\tilde{E}_0^2\Delta\sin(2kx)$ with $k = 2\pi/L_x$ and
$\tilde{g}\tilde{E}_0^2\Delta = 1.\times 10^{-2}$. Our initial
conditions are again such that one soliton is present.}
\label{lower_om15_sin2}
\end{center}
\end{figure}

The behavior of the time-dependent voltages at different points
along the junction can be understood from the sinusoidal
junction-cavity coupling.  For every point on the junction, there is
a corresponding point separated by $L_x/2$ which experiences an
equal and opposite coupling to the cavity. If the coupling has a
spatial dependence $\sin(2\pi x/L_x)$, there are two nodal points
along the junction where the coupling is zero, and two points where
the coupling has maximum amplitude but is 180$^o$ out of phase.  At
these two maximum points, the cavity-soliton interaction at any
given time is equal in magnitude but opposite in sign.  Because of
this feature, the time-dependent voltages at these two points should
have the same wave form but should be 180$^o$ out of phase, as is
seen in Figs.\ \ref{phase60_120} (a) and (b).

This picture also explains why the time-dependent
 voltages along the junction are at all points equal in pairs but
180$^o$ out of phase.   The different pairs have different voltage
wave-forms because the coupling amplitude between each mini-junction
and the cavity varies with the spatial dependence $\sin(kx)$.
Despite the different wave-forms, we have verified that the {\em
time-averaged} voltage difference is the same at each point along
the junction. This is consistent with the picture that this voltage
is produced by a soliton which passes each point along the junction
with the same average frequency.

Further evidence of a strong soliton-cavity coupling can be seen by comparing
Figs.\ \ref{lower_om75_g02} and \ref{lower_om15_sin2}.  Fig.\
\ref{lower_om75_g02}, as noted above, shows the soliton branch of the IV curve for
$\tilde{\Omega} = 0.75$, $\tilde{g}\tilde{E}_0^2\Delta = 10^{-2}$,
and a coupling varying spatially as $\sin(2\pi x/L_x)$. In the
second, we assume the same coupling strength of
$\tilde{g}\tilde{E}_0^2\Delta = 10^{-2}$ but a frequency
$\tilde{\Omega} = 1.5$ and a spatial dependence of $\sin(4\pi
x/L_x)$.  In both cases, there are steps at approximately the same
value of $\langle V \rangle$, namely $\langle V\rangle/(NRI_c) \sim
0.33 \sim \pi/10$.   These figures show that simultaneously doubling
the cavity frequency and halving the coupling periodicity leaves the
step height unchanged.  We interpret this behavior as showing that,
in both cases, the soliton is locked to the cavity mode so that it
moves by a distance equal to the wavelength of the mode per mode
cycle.

We have also computed $q_r(\tau)$ for $J_z/J_c= 0.05$ and the
parameters of Fig.\ \ref{lower_om15_sin2} (not shown).  We find that
this amplitude is much smaller than that shown in Fig.\ 11 for the step of
Fig.\ \ref{lower_om75_g02}. A simple qualitative explanation for
this behavior is the following.  The soliton has the same spatial
width in both cases, but the junction-cavity coupling varies
spatially more rapidly in \ref{lower_om15_sin2} than in Fig.\
\ref{lower_om75_g02}.  Since the junction-cavity coupling thus
varies substantially over the width of the soliton in Fig.\
\ref{lower_om15_sin2}, it has a smaller effect than in Fig.\
\ref{lower_om75_g02}.  Thus, we expect a much weaker step-like
feature in the IV characteristic of Fig.\ \ref{lower_om15_sin2} than
in Fig.\ \ref{lower_om75_g02}, as is indeed observed numerically.

\section{Discussion}

\subsection{Qualitative Argument for Soliton Steps}

The locking of the moving soliton to the mode of the resonant cavity
can be accounted for by a simple analytical argument.  The argument
starts from the equations of motion (\ref{eq:qj}) and (\ref{eq:qc}).
We assume solutions of the form
\begin{eqnarray}
\phi(x, t) = \phi_0 + k(x - vt) + \phi_1\sin[k(x-vt)] \label{eq:s1}
\\
q_r(t) = \mathrm{Re}(q_0e^{i\omega t}) \label{eq:s2}
\end{eqnarray}
This solution is suggested by the fact that, if $g^\prime = 0$, the
solution $\phi(x, t)$ is rigorously of the form $\phi(x, t) = \phi(x
- vt)$, where $v$ is the soliton velocity.  We now substitute these
assumed solutions into eqs.\ (\ref{eq:qj}) and (\ref{eq:qc}), carry
out the indicated derivatives, and use the standard expansion
\begin{equation}
\sin\left\{\phi_0+k(x-vt)+ \phi_1\sin[k(x-vt)]\right\} =  \sum_{n =
-\infty}^\infty J_n(\phi_1)\sin[\phi_0+k(x-vt)+nk(x-vt)],
\label{eq:bessel}
\end{equation}
where $J_n(\phi_1)$ is the n$^{th}$ order Bessel function.  In eq.\
(\ref{eq:qc}), we use the expression $\phi_{tt} =
-\omega^2\phi_1\sin[k(x-vt)]$, where $\omega = vk$, and carry out
the integral to obtain
\begin{equation}
\ddot{q}_r + \frac{\Omega}{Q_c}\dot{q}_r + \Omega^2q_r =
-\frac{1}{2}\omega^2L_x\phi_1\cos(\omega t). \label{eq:qc1}
\end{equation}
Since this equation is linear, the driven solution for $q_r$ is
simply
\begin{equation}
q_r(t) = -\frac{1}{2}Kg^\prime\omega^2L_x\phi_1\mathrm{Re}
\left(\frac{e^{i\omega t}}{\Omega^2 + i\Omega\omega/Q_c -
\omega^2}\right). \label{eq:qc2}
\end{equation}

Solution (\ref{eq:qc2}) can now be substituted, along with the
Bessel function expansion, eq.\ (\ref{eq:bessel}), back into the
other equation of motion, eq.\ (\ref{eq:qj}).   Next, we assume that
the amplitude $\phi_1$ is small, and expand the Bessel functions in
powers of $\phi_1$.  Finally, we retain only the lowest Fourier
components, namely those involving a constant term, and the
functions $\sin[k(x-vt)]$ and $\cos[k(x-vt)]$.  A similar approach
has been used previously in Refs.\ \cite{helm1} and \cite{preis} to
treat intrinsic Josephson junctions to optical phonons in
high-temperature superconductors.  Setting the coefficients of each
of these terms separately equal to zero in the expanded version of
eq.\ (\ref{eq:qj}), we finally obtain the following three equations
for $\phi_0$, $\phi_1$, and $(J/J_c)$:
\begin{equation}
\frac{1}{\lambda_J^2}\frac{J}{J_c}-\frac{\omega_p\omega}{Q_J\bar{c}^2}
+\frac{1}{\lambda_J^2}\frac{\phi_1}{2}\sin\phi_0  =  0;
\label{eq:res1}
\end{equation}
\begin{equation}
 \frac{\omega_p\omega}{Q_J\bar{c}^2}\phi_1 +
\frac{1}{\lambda_J^2}\sin\phi_0 +
\frac{K^{\prime\prime}(\Omega\omega/Q_c)}{4D} = 0 \label{eq:res2};
\end{equation}
\begin{equation}
 -k^2(1 - \frac{v^2}{\bar{c}^2})\phi_1 +
\frac{1}{\lambda_J^2}\cos\phi_0 +
\frac{K^{\prime\prime}(\Omega^2-\omega^2)}{4D} = 0, \label{eq:res3}
\end{equation}
where $\omega = vk$, $k = 2\pi/L_x$, $K^{\prime\prime} = (\hbar^2c^2
g^\prime)^2\omega^4L_x/[16\pi d (e^*)^2M]$ and
\begin{equation}
D = (\Omega^2 - \omega^2)^2 + \Omega^2\omega^2/Q_C^2 \label{eq:res4}
\end{equation}
is a resonant denominator\cite{note3}.

Eqs.\ (\ref{eq:res2}) and (\ref{eq:res3}) are readily solved for
$\phi_0$ and $\phi_1$ in terms of the soliton velocity $v$, and the
result substituted into eq.\ (\ref{eq:res1}).  The resulting eq.\
(\ref{eq:res1}) expresses the current $J$ in terms of the soliton
velocity $v$, or equivalently, the time-averaged voltage across the
long junction, $\langle V \rangle_ t = 2\pi \hbar v/(2eL_x)$. As is
suggested by the resonant form of eqs.\ (\ref{eq:res2}) and
(\ref{eq:res3}), this IV characteristic has a peak when $\omega \sim
\Omega$, corresponding to the voltage plateaus observed in our
numerical results.

This same approach also shows, through eq.\ (\ref{eq:qc2}), that
there is a peak in the amplitude $q_r$ of the cavity mode when the
same condition ($\omega = \Omega$) is satisfied.  Once again, this
peak in the amplitude is observed in our numerical simulations.  For
example, Fig.\ 11 shows the large amplitude of $q_r$ at a voltage
satisfying the resonance condition; the amplitude of $q_r$ at other
voltages is much smaller.

\subsection{Comparison to Predictions of Other Models}

It is useful to compare our model and results for the coupled
soliton-cavity system to those of other workers.  For the case of a
{\em uniform} coupling, our results are formally analogous to those
obtained in Refs.\ \cite{helm1} and \cite{helm2}.  These workers
consider the response of {\em intrinsic} Josephson junctions in a
high-T$_c$ superconductor coupled to an {\em optical phonon} mode
within the junction.  For spatially varying coupling, our numerical
results for the locking of a moving fluxon to a cavity mode somewhat
resemble those of Refs.\ \cite{preis} and \cite{kleiner3},
obtained for a coupled fluxon/optical phonon system in an intrinsic
Josephson junction, though the equations describing the two systems
are not identical.  However, there is a significant difference in
the physics.  The anomalies found in Refs.\ \cite{preis} and
\cite{kleiner3}: in these systems, the quantities which play the
role of cavity modes are the optical phonons, which are intrinsic to
the junctions themselves.  By contrast, our cavity modes are assumed
to arise from some cavity extrinsic to the junctions.

\subsection{Possible Realizations of Sinusoidal Coupling}

Finally, we briefly discuss the type of electromagnetic mode which
could produce the sinusoidal coupling we use.  The electric field of
this mode has a non-zero curl which varies sinusoidally with
position, i. e., it has a sinusoidally varying {\em magnetic} field.
This type of mode should be readily achievable in a real cavity. The
voltage steps should be achievable so long as the phase velocity in
the cavity is smaller than the limiting soliton velocity $\bar{c}$.
Another way to produce this type of coupling would be actually to
prepare a long Josephson junction in the shape of a ring, and then
to use a cavity mode with a {\em spatially uniform} magnetic field.
In this case, the junction occupies the ribbon-like region between
two circular rings of superconductor; if the planes of the rings are
parallel to the z axis, then the magnetic field of the cavity mode
should be uniform and parallel to one of the ring diameters.  This
will produce a flux through the long junction which varies
sinusoidally around the ring, as in our model.  Measurements using a
ring geometry, and a {\em static} uniform magnetic field, have
recently been reported, in another context, by Wallraff {\it et
al.}\cite{wallraff}, who also show a schematic picture of this
geometry.

\section{Summary}

In this paper, we have described a model for a long underdamped
Josephson junction interacting with a single-mode electromagnetic
cavity.  In our model, we have assumed a capacitive interaction
between the junction currents and cavity mode, but we consider both
a uniform coupling and one which varies spatially along the junction
length. If no soliton (i. e. no fluxon) is present, the junction
behaves very much like a small Josephson junction\cite{almaas}.  In
particular, there are SIRS's just as in a small junction, which
occur at the voltages expected for a small junction.

If a soliton is present, and the junction-cavity coupling is
position-independent, then we find that there are {\em no} SIRS's.
The absence of SIRS's in this case is easily understood: since the
soliton-cavity interaction energy is independent of the soliton's
position, the cavity exerts no force on the soliton. If, however,
the coupling varies sinusoidally with position, we find step-like
structure in the IV characteristics, which arise from the
junction-cavity coupling. These structures arise from the locking of
the soliton to the cavity mode, so that the soliton travels a
distance equal to one wavelength of the coupling interaction during
one cycle of the cavity mode, or equivalently, the average soliton
velocity equals the phase velocity $\omega/k$ of the cavity mode. We
find clear evidence of the locking between the cavity mode and the
junction through the time-dependent voltages across the junction.
Everywhere along the junction, these time-dependent voltages have
the same period of oscillation as the cavity mode. Moreover, the
voltage differences of points separated by half a wavelength have
exactly the same wave form, but are 180$^o$ out of phase, showing
that the soliton travels one wavelength per cycle of the cavity
mode.

We have also presented a simple qualitative argument which explains
both the positions of the self-induced voltage steps and the
occurrence of a peak in the radiated energy on the steps.  This
argument agrees well with our numerical results in the limit of weak
junction-cavity coupling.

If the junction-cavity coupling is strong, this simple argument does
not give the position of the anomalies in the IV characteristics.
Instead, the voltage of the step-like structure is shifted
substantially {\em down}, and the frequency of the cavity mode is
strongly red-shifted. Nonetheless, the cavity is still locked to the
soliton motion.

The voltage on the steps is not absolutely constant, but varies
slightly with current.  When the current lies on a voltage step, the
cavity is strongly excited, with a large time-averaged squared
amplitude $q_r^2$.  If the current does not lie on a step, cavity
and Josephson junction are not locked, and $q_r$ is much smaller (at
least by an order of magnitude in all our numerical runs).  This
behavior is once again in agreement with our simple analytical model
of the previous section.

The present model could be modified to apply to a {\em stack} of
long junctions coupled to a single-mode cavity.  In the absence of a
cavity, it has been known for some time that very anisotropic
high-T$_c$ cuprate superconductors, such as \bscco, behave like a
stack of underdamped Josephson junctions\cite{kleiner1}. Coupling
such a stack to a cavity is of great interest because it may provide
a means for phase-locking these junctions, and hence providing a
coherent source of sub-THz radiation. The dynamics of a stack of
long junctions {\em without} a cavity has previously been modeled by
Sakai {\it et al} \cite{sakai}.   A stack of long junctions coupled
to internal phonon modes within the junctions has been modeled by
Preis {\it et al}\cite{preis}. The present work suggests a means of
modeling the coupling of a stack of junctions to {\em the same}
electromagnetic cavity.

\section{Acknowledgments}

This work was supported by NSF Grant DMR04-13395.
The calculations were carried out using the Ohio Supercomputer
facilities. We are very grateful for valuable conversations with
Prof. B. R. Trees and Dr. E. Almaas.

\section{Appendix: Alternate Derivation of
Lagrangian for Long Junction
Coupled to a Cavity.}

\vspace{0.1in}

In this Appendix, we present an alternate derivation of the
cavity-junction Lagrangian obtained in Section II.  The derivation
here is more general, in that we consider coupling to both the
electric and magnetic fields of the cavity.

We write the total Lagrangian as
\begin{equation}
L = K_1 - U_1.
\end{equation}
The kinetic energy, K, is written as
\begin{equation}
K_1 = \int d^3x \frac{\epsilon({\bf x}){\bf E}\cdot {\bf E}}{8\pi},
\end{equation}
where $\epsilon({\bf x})$ is the (possibly position-dependent) dielectric
function, and ${\bf E}({\bf x})$ is the electric field.  We assume
that the electric field is the sum of two parts: that due to the junction,
which we denote ${\bf E}_{junc}$, and the part due to cavity mode,
which we write as ${\bf E}_{cav}$.  The junction field takes the form
\begin{equation}
{\bf E}_{junc} = \frac{\hbar}{2ed}\phi_t{\bf \hat{z}}.
\end{equation}
Then $K_1$ takes the form
\begin{eqnarray}
K_1  =  L_y\int dx\frac{\epsilon\hbar^2\phi_t^2}{8\pi(2e)^2d} +
\frac{1}{8\pi}\int d^3x\epsilon({\bf x})({\bf E}_{cav}\cdot{\bf E}_{cav})
+ \nonumber \\
+  \frac{1}{4\pi}\int d^3x\epsilon({\bf x}){\bf E}_{cav}({\bf x})\cdot
\frac{\hbar}{2ed}\phi_t{\bf \hat{z}} \nonumber .
\end{eqnarray}
Here the volume integral is taken over the cavity, and we are assuming
a geometry such that the junction is contained within the cavity.

The potential energy $U_1$ is the sum of two terms.  The first is the
Josephson energy,
\begin{equation}
U_{1,J} = -L_y\int dx \frac{\hbar J_c}{2e}\cos\phi.
\end{equation}
The other part is the energy of the magnetic field.  This may be written
\begin{equation}
U_{1,mag} =\int d^3x\frac{{\bf B}\cdot{\bf B}}{8\pi}.
\end{equation}
Once again, the magnetic field may be written
\begin{equation}
{\bf B} = {\bf B}_{junc} + {\bf B}_{cav},
\end{equation}
where ${\bf B}_{junc}$ and ${\bf B}_{cav}$ are the fields due to the
junction and cavity.   As in the text, we again assume that $J_u = J_\ell$,
so that
\begin{equation}
{\bf B}_{junc} = \frac{\Phi_0\phi_x}{2\pi d}{\bf \hat{y}}
\end{equation}
Then the magnetic field energy takes the form
\begin{eqnarray}
U_{1,mag} = L_y\int dx \frac{\Phi_0^2\phi_x^2}{32\pi^3 d}
+\frac{1}{8\pi}\int d^3x {\bf B}_{cav}\cdot{\bf B}_{cav} +
\nonumber \\
+ L_y\int dx \frac{1}{4\pi}\frac{\Phi_0\phi_x}{2\pi}\hat{y}\cdot
{\bf B}_{cav}({\bf x}) \nonumber.
\end{eqnarray}

To make further progress, we introduce operators describing the
fields ${\bf E}_{cav}$ and ${\bf B}_{cav}$.  Both may be expressed
in terms of the operator for the cavity vector potential:
\begin{equation}
{\bf A}_{cav}({\bf x}, t) = \left(\frac{hc^2}{\Omega}\right)^{1/2}
[a(t) + a^\dag(t)]{\bf E}({\bf x}).
\end{equation}
Here $a$ and $a^\dag$ are the annihilation and creation operators for
the cavity mode, which satisfy the usual Bose commutation relations
\begin{equation}
[a, a^\dag] = 1.
\end{equation}
${\bf E}({\bf x})$ is proportional to the position-dependent
electric field of the cavity mode; its normalization is given below.
$\Omega$ is the frequency of the cavity mode.  In terms of this
operator, the electric field operator ${\bf E}_{cav}({\bf x}, t)$
is given by
\begin{equation}
{\bf E}_{cav}({\bf x}, t) =
-\frac{1}{c}\frac{\partial{\bf A}_{cav}}{\partial t}
= -i(h\Omega)^{1/2}(a - a^\dag){\bf E}({\bf x})
\end{equation}
and the magnetic field operator ${\bf B}_{cav}({\bf x}, t)$ is
\begin{equation}
{\bf B}_{cav}({\bf x}, t) = {\bf \nabla}\times{\bf A}_{cav}
= \left(\frac{hc^2}{\Omega}\right)^{1/2}[a + a^\dag]{\bf \nabla}\times
{\bf E}({\bf x}).
\end{equation}
The operator describing the total energy in the cavity is
\begin{equation}
W_{cav} = \int d^3x \frac{{\bf E}_{cav}\cdot{\bf E}_{cav} + {\bf B}_{cav}
\cdot{\bf B}_{cav}}{8\pi},
\end{equation}
We calculate the ensemble average of this operator, using the results
$\langle a^\dag a \rangle = n$; $\langle aa^\dag\rangle = n+1$;
$\langle aa\rangle = \langle a^\dag a^\dag \rangle =0$, with the result
\begin{equation}
\langle W_{cav}\rangle = \frac{1}{2}\left(n+\frac{1}{2}\right)\hbar\Omega
\int \left[|{\bf E}|^2 + \frac{c^2}{\Omega^2}|{\bf \nabla}\times {\bf E}|^2
\right]d^3x.
\end{equation}
In order for this energy to equal $\hbar\Omega(n+\frac{1}{2})$, we
require that the function ${\bf E}({\bf x})$ be normalized so that
\begin{equation}
\int \left[|{\bf E}|^2 + \frac{c^2}{\Omega^2}|{\bf \nabla}\times
{\bf E}|^2\right]d^3x = 2.
\end{equation}
Also, we assume that the cavity energy is equally distributed between
the electric and magnetic fields.  This implies that
\begin{equation}
\int |{\bf E}|^2d^3x = \frac{c^2}{\Omega^2}\int|{\bf \nabla}
\times {\bf E}|^2 d^3x = 1.
\end{equation}

Having obtained the operator forms for the cavity electric and magnetic
fields, we are now in a position to derive the electric and magnetic
parts of the junction-cavity coupling.  After some
algebra, the electric field part of the coupling Lagrangian may be written
\begin{equation}
L_{coup,E}= \frac{1}{4\pi}L_y\int dx \frac{\hbar}{2e}\phi_t[-i(h\Omega)^{1/2}
(a - a^\dag)]E_z({\bf x}).
\end{equation}
Similarly, the magnetic field part of the coupling Lagrangian may be written
\begin{equation}
L_{coup,B} = \frac{1}{4\pi}L_y\frac{\Phi_0}{2\pi}(hc^2/\Omega)^{1/2}
\int dx (a + a^\dag)\phi_x({\bf \nabla} \times {\bf E}({\bf x}))_y.
\end{equation}

To make contact with the notation in the main part of this paper,
we introduce position and momentum operators for the cavity mode, by
\begin{equation}
p_r = \left(\frac{M\hbar\Omega}{2}\right)^{1/2}i(a^\dag-a)
\end{equation}
and
\begin{equation}
q_r = \left(\frac{\hbar}{2M\Omega}\right)^{1/2}(a+a^\dag).
\end{equation}
These operators have the standard canonical commutation relations
\begin{equation}
[p_r, q_r] = -i\hbar.
\end{equation}
In terms of these operators, we may write the Hamiltonian
for the cavity alone as
\begin{equation}
H_{cav} = \frac{1}{8\pi}\int\left({\bf E}_{cav}\cdot{\bf E}_{cav}
+ {\bf B}_{cav}\cdot{\bf B}_{cav}\right)
= \frac{p_r^2}{2M} + \frac{Kq_r^2}{2},
\label{eq:hcav}
\end{equation}
where we define the "spring constant"
\begin{equation}
K = M\Omega^2.
\end{equation}
From $H_{cav}$ we may infer the corresponding
cavity Lagrangian using $q_r = p_r/M$, with the result
\begin{equation}
L_{cav} = \frac{M\dot{q}_r^2}{2} - \frac{1}{2}Kq_r^2.
\label{eq:lcav}
\end{equation}
It is readily verified that classical Hamiltonian
equations of motion resulting from the Hamiltonian
(\ref{eq:hcav}) are equivalent to the classical Lagrange equation of
motion obtained from the Lagrangian (\ref{eq:lcav});
both lead to
\begin{equation}
\ddot{q}_r - \Omega^2 q_r = 0.
\end{equation}

Finally, we can express the coupling Lagrangians in terms of
the variables introduced above.  First $L_{coup,E}$ takes the
form
\begin{equation}
L_{coup,E} = L_y\int dx {\cal L}_{coup,E}(x),
\end{equation}
where
\begin{equation}
{\cal L}_{coup,E}(x) = -g(x)\dot{q}_r\phi_t,
\end{equation}
and
\begin{equation}
g(x) = -\epsilon\sqrt{\frac{M}{4\pi}}\frac{\hbar}{2e}
{\bf E}(x)\cdot{\bf \hat{z}}.
\end{equation}
Similarly, $L_{coup,B}$ takes the form
\begin{equation}
L_{coup,B} = -L_y\int dx {\cal L}_{coup,B}(x),
\end{equation}
where
\begin{equation}
{\cal L}_{coup,B}(x) = -g_B(x)q_r\phi_x,
\end{equation}
and
\begin{equation}
g_B(x) = \sqrt{\frac{Mc^2}{16\pi^3}}\Phi_0{\bf \nabla}\times
{\bf E}\cdot{\bf \hat{y}}.
\end{equation}

There is one additional term in the Lagrangian which also represents
a coupling between the cavity and the junction.  This is the term
\begin{equation}
L_{d} = L_y\int dx \frac{d}{8\pi}(\epsilon - 1){\bf E}_{cav}\cdot{\bf E}_{cav}.
\end{equation}
But this term does not couple the cavity variables to those of the junction.
Instead, its only effect will be to produce a slight shift in the cavity
resonance frequency.  We have therefore not included this term in our
calculations described in the body of the paper.


\begin{thebibliography}{99}
\bibitem{note3} In obtaining these equations, we have written
terms of the form $\sin(kx)\cos(\omega t) =
\frac{1}{2}[\sin(kx-\omega t) + \sin(kx+\omega t)]$ and
$\sin(kx)\sin(\omega t) = \frac{1}{2}[\cos(kx-\omega t) -
\cos(kx+\omega t)]$ and neglected the "counter-rotating" terms
$\cos(kx+\omega t)$ and $\sin(kx+\omega t)$.
\end{thebibliography}
\end{document}